\documentclass[12pt,preprint]{aastex}

\def\lapprox{\;\rlap{\lower 2.5pt                       
             \hbox{$\sim$}}\raise 1.5pt\hbox{$<$}\;}

\shorttitle{Interstellar Krypton Abundance Variations}
\shortauthors{Cartledge et al.}

\begin{document}

\title{Interstellar Krypton Abundances: The Detection of Kiloparsec-scale
Differences in Galactic Nucleosynthetic History\footnote{Based on observations
with the NASA/ESA {\it Hubble Space Telescope (HST)} and the NASA-CNES-CSA {\it
Far-Ultraviolet Spectroscopic Explorer (FUSE)}. {\it HST} spectra were obtained
at the Space Telescope Science Institute, which is operated by the Association
of Universities for Research in Astronomy, Inc. under NASA contract No. NAS
5-26555; {\it FUSE} is operated for NASA by the John Hopkins University under
NASA contract NAS-32985.}}

\author{Stefan I.B. Cartledge}
\affil{Dept. of Physics and Astronomy, Louisiana State University, Baton Rouge,
    LA 70803}
\email{scartled@gmail.com}

\author{J. T. Lauroesch}
\affil{Department of Physics and Astronomy, University of Louisville,
    Louisville, KY 40292}
\email{jtlaur01@louisville.edu}

\author{David M. Meyer}
\affil{Department of Physics and Astronomy, Northwestern University, Evanston,
    IL 60208}
\email{davemeyer@northwestern.edu}

\author{Ulysses J. Sofia}
\affil{Department of Astronomy, Whitman College, Walla Walla, WA 99362}
\email{sofiauj@whitman.edu}

\and

\author{Geoffrey C. Clayton}
\affil{Dept. of Physics and Astronomy, Louisiana State University, Baton Rouge,
    LA 70803}
\email{gclayton@fenway.phys.lsu.edu}

\begin{abstract}
We present an analysis of \ion{Kr}{1} $\lambda$1236 line measurements from 50
sight lines in the {\it Hubble Space Telescope} Space Telescope Imaging
Spectrograph and Goddard High Resolution Spectrograph data archives that have
sufficiently high resolution and signal-to-noise ratio to permit reliable
krypton-to-hydrogen abundance ratio determinations. The distribution of Kr/H
ratios in this sample is consistent with a single value for the ISM within 5900
pc of the Sun, log$_{10}$(Kr/H) = $-9.02\pm0.02$, apart from a rough annulus
from between $\sim$600 and 2500 pc distant. The Kr/H ratio toward stars within
this annulus is elevated by approximately 0.11 dex, similar to previously noted
elevations of O/H and Cu/H gas-phase abundances beyond $\sim$800 pc. A
significant drop in the gas-phase N/O ratio in the same region suggests that
this is an artifact of nucleosynthetic history. Since the physical scale of the
annulus' inner edge is comparable to the radius of the Gould Belt and the outer
limit of heliocentric distances where the D/H abundance ratio is highly
variable, these phenomena may be related to the Gould Belt's origins.
\end{abstract}


\section{Introduction}
Where krypton is detectable, its interstellar abundance has the capacity to be
a singularly reliable gauge of the degree of nucleosynthesis in a given region
of the interstellar medium (ISM). Krypton is a noble gas whose outer electron
shell is spherically symmetric. Consequently, it is not prone to forming either
mechanical or chemical bonds in diffuse interstellar clouds where, given its
ionisation potential relative to that of hydrogen (13.999 eV versus 13.598 eV),
it should be predominantly in neutral form. These properties imply that a
constant krypton to hydrogen interstellar abundance ratio is synonymous with a
well-mixed ISM on the length scale over which the gas is sampled. Conversely,
any significant departures from an established large-scale mean represent
nucleosynthetic artifacts that mixing has not yet erased. Identifying such
artifacts is crucial to understanding topics as varied as the details of dust
composition and the large-scale processes of galactic chemical evolution. In
particular, analyses of abundance departures from interstellar means constrain
the observed efficacy of mixing at different Galactic length scales, the
influence of isolated and/or prolonged but localized star formation events on
elemental abundance ratios, the yields of supernovae, and the amount and
composition of material available for dust formation. Because it is undepleted
in the diffuse ISM, elemental abundance ratios involving krypton can be very
sensitive probes for these fields of study.

Krypton to hydrogen abundance ratios have been well-studied within a kiloparsec
of the Sun. The krypton line at 1235.838 {\AA} was first detected using the
Goddard High Resolution Spectrograph (GHRS) aboard {\it Hubble Space Telescope
(HST)} during observations of $\zeta$ Oph \citep{car91}. A comprehensive study
of GHRS detections of \ion{Kr}{1} $\lambda$1236 later demonstrated that the
Kr/H abundance ratio within 500 pc of the Sun is constant at the level of
log$_{10}$(Kr/H) = $-9.02\pm0.02$ \citep{car97}. This conclusion was consistent
with GHRS results from oxygen \citep{mey98} and nitrogen \citep{mey97}, two of
the most abundant elements in the ISM, which also exhibited a constant local
abundance relative to hydrogen. Moreover, since the interstellar krypton
abundance ratio is roughly 50\% of the Solar value ($-8.72\pm0.08$;
\citealt{lod03}), this study established that the deficit in the interstellar
krypton abundance with respect to its Solar value was similar to the deficits
for other elements (e.g., carbon, oxygen, and nitrogen), using the then-accepted Solar abundances. Myriad spectral observations using the Space Telescope Imaging Spectrograph (STIS) and revisions both to oscillator strengths and Solar abundances have lessened the similarity between the interstellar abundance deficits for these elements; nevertheless the uniformity of their elemental abundances in the local ISM has only {\it gained} further support from the new data. Indeed, since krypton does not readily deplete into grains or form molecules, the constancy of its abundance relative to hydrogen, independent of molecular hydrogen fraction [$f$(H$_2$)] or mean hydrogen sight line density [${\langle}n_{\rm H}\rangle$], has been used to establish that the diffuse ISM is well-mixed on lengthscales of a few hundred parsecs \citep{car03,car06}. This result is independently supported by other analyses, specifically some involving the composition of stellar atmospheres \citep{red03} and pre-solar meteoritic dust grains \citep{nit05}. As a result, it can be concluded that significant deviations from the mean interstellar Kr/H abundance ratio identify sites or regions within the Galaxy where the nucleosynthetic history differs from that of the local ($d \lapprox$ 500-700 pc) ISM.

In order to apply this tool to exploring the limits of this well-mixed
interstellar region and to probe for variations in the Kr/H abundance ratio
based on physical conditions not found locally, it was necessary to search for
krypton along a wider variety of paths than were previously studied. Thus, we
have plumbed the depths of the MAST Archive at STScI for detections of
\ion{Kr}{1} $\lambda$1236, measuring the krypton abundance for each sight line
with a significant absorption feature. In this paper, we discuss the relevant
GHRS, STIS, and {\it Far Ultraviolet Spectroscopic Explorer (FUSE)}
observations (\S~\ref{observations}), then examine the uniformity of the Kr/H
abundance ratio with respect to key sight line properties
(\S~\ref{krypton_distribution}). Finally, in \S~\ref{distance_effect},
variations in Kr/H, O/H, O/Kr, N/O, and Cu/H are compared, in search of
reinforcement for the identified Kr/H trends, and possible explanations.

\section{Observations and Abundance Measurements}
\label{observations}
The study of interstellar krypton abundances using GHRS initially established
our knowledge of the uniformity of the Kr/H ratio within 500 pc of the Sun \citep{car97}. When
STIS data were added to the analysis, the case for a singular interstellar Kr/H
abundance ratio was broadened to include 26 sight lines covering a range of
nearly five orders of magnitude in $f$(H$_2$) and over two orders of magnitude
in ${\langle}n_{\rm H}\rangle$, as only two exceptions to the remarkably narrow
distribution around log$_{10}$(Kr/H) = $-9.02\pm0.02$ were identified
\citep{car03}. Since the majority of the combined GHRS and STIS sample were
relatively short sight lines ($<$ 600 pc) and the two exceptions (HD 116852 and
HD 152590) were the only paths to penetrate neighboring spiral arms, these
deviations were thought to be related to this circumstance. Seeking to
investigate this hypothesis, we proposed to use STIS to observe eight Galactic
OB stars (GO9855), at distances up to 4.8 kpc, that intersect neighboring
spiral arms. The targets included both HD 116852 and HD 152590, so that the
previous krypton measurements could be compared with results derived from new
higher signal-to-noise ratio spectra. Each observation was designed to generate
spectra with S/N = 50 per resolution element for the E140H echelle grating
centered at 1271 {\AA} using the {0.2\arcsec$\times$0.9\arcsec} aperture. The
equivalent width of an unsaturated feature detected at 5$\sigma$ would thus be
1.1 m{\AA}, permitting the measurement of log$_{10}$(Kr/H) values as small as
$-9.60$; this limit is a factor of three smaller than the smallest ratio
reported to that date.

Unfortunately, STIS ceased to operate in mid-2004, after only four of our
planned targets were observed. Although we were now able to re-examine both
previously observed targets, only two new sight lines, one each directed toward
and away from the Galactic centre, were added. To achieve the stated goals of
GO9855, we were forced to search the STIS archive for all observations that
might possess unreported detections of krypton in either the 1164.8672 or
1235.838 {\AA} resonance lines. Since this search was well-matched to our
archival proposal to investigate the abundances of rare elements (AR10643;
\citealt{lau08}), particularly those produced by slow and rapid neutron capture
processes, the searches were conducted concurrently. The relevant search
criteria were sufficient spectral resolution to distinguish components from
different spiral arms and limit contamination from unresolved saturation (no
datasets with resolution less than $\lambda/\Delta\lambda$ = 45000 were used),
and wavelength coverage that included one or both of the krypton resonance
lines previously mentioned. Among the archival datasets, 35 sight lines exhibit
some evidence of a krypton absorption feature associated with \ion{Kr}{1}
$\lambda$1236; observational details for the 26 sight lines with reliable
detections and the four GO9855 targets are given in Table~\ref{datalist}. The
table includes specifications for all GHRS, STIS, and {\it FUSE} observations
of these stars upon which the abundance measurements are based. Unfortunately,
the poor signal-to-noise ratios for spectral orders covering 1164.8672 {\AA} do
not permit reliable column density determinations from these data. Please note
that seven sight lines with marginal krypton detections (HD 3827, HD 93205, HD
93222, HD 168941, HD 202347, HD 218915, and CPD $-$59 2603) have been rejected
because in each case the absorption feature's equivalent width was less than a
factor of 4 greater than the associated uncertainty. HD 36408B and HD 121968
have better than 5$\sigma$ features, but these are omitted for the reasons
outlined below.

The spectral type for HD 36408 B, B7IV, is unsuitable for the direct
measurement of an interstellar atomic hydrogen abundance, so this quantity must
instead be gauged using the well-known empirical linear relationship between
$E(\bv)$ and $N$(H) (e.g., see \citealt{spi78}). Using the \citet{rac02}
calibration indicates an oxygen abundance of log$_{10}$(O/H) = $-3.17\pm0.30$,
about 0.24 dex above the interstellar mean for low-density sight lines
\citep{car04}. The gas-phase abundances of other elements, notably \ion{Mg}{2},
\ion{P}{2}, \ion{Ni}{2}, \ion{Cu}{2}, \ion{Ge}{2}, and especially \ion{Kr}{1},
range from 0.30 dex to 0.60 dex higher than their respective low-density
interstellar means, while \ion{Mn}{2} is roughly at the interstellar value (for
interstellar abundance references at log$_{10}{\langle}n_{\rm H}\rangle \approx
-0.11$, see \citealt{car06}). Due to the lack of $N$(H) and the unusual
depletion pattern along this sight line compared with sight lines analyzed by
\citet{car06}, we identify this path as a subject for future study but omit it
from the current analysis.

We also identify HD 121968 as a sight line whose ambiguous krypton abundance
warrants further study. A strong absorption component is evident near
$v_{heliocentric}= -10$km/s in line profiles of \ion{O}{1} $\lambda$1355,
\ion{Si}{1} $\lambda$1425, \ion{P}{2} $\lambda$1301, all \ion{Mn}{2} lines,
\ion{Cu}{2} $\lambda$1358, \ion{Ga}{2} $\lambda$1414, \ion{Ge}{2}
$\lambda$1237, all \ion{C}{1} lines, and \ion{Cl}{1} $\lambda$1347. The
\ion{Kr}{1} $\lambda$1235 profile contains a similar feature at the same
helocentric velocity; however, an absorption spike of equal strength is
situated 5 km/s to the blue. In none of the other line profiles mentioned above
does a comparable component appear, which would make this sight line unique
among those studied to date if the krypton profile were accepted without
further consideration. Several nickel lines are the only profiles to exhibit
strong absorption at velocities removed from the central component; the feature
most closely approximating the second krypton absorption spike is at
$v_{heliocentric}\approx -18$km/s with nearly 75\% depth relative to the main
component. However, this feature is merely one of the strongest in a very
complex profile that bears very little resemblance to the line profile of
krypton or any other element. Lastly, there is no blueward component to match
the hint of absorption that corresponds to the main absorption feature in the
the \ion{Kr}{1} $\lambda$1164 profile, although this spectral order is
compromised by noise spikes comparable to the profile depth. Consequently, we
conjecture that the apparent \ion{Kr}{1} $\lambda$1235 absorption profile is
contaminated by noise, and omit the HD 121968 sight line from further analysis.

Adding 30 new krypton gas-phase abundance measurements to previous
detections, there now exists a database of 56 Galactic sight lines with
published krypton column densities. Unfortunately, for three of these paths
there are no observations at short enough wavelengths to assess the amount of
molecular hydrogen intersected by the line of sight. Toward HD 37021 and HD
37061, however, \citet{car01} argued that the low \ion{Cl}{1} abundance toward
each star implies that these sight lines do not have large H$_2$ column
densities. In the current sample, HD 208947 lacks this evidence for a small
$f$(H$_2$) but its krypton abundance is near the low end of the distribution
and we include it provisionally in the Kr/H abundance ratio analysis. However,
six stars whose spectra exhibit krypton absorption features also have
interstellar Lyman-$\alpha$ absorption profiles that are contaminated by
significant stellar absorption, hence the sample of Kr/H abundance ratios
contain values for only 50 of the 56 sight lines with krypton. The krypton
abundance for each path was determined using the methods established by
\citet{car01}, with a slight alteration to the calibration procedure. The new
spectra were calibrated using the standard STSDAS STIS data reduction routine
CALSTIS (v.2.13), with the two-dimensional scattered-light correction algorithm
SC2DCORR applied automatically. Although this procedure produces slightly
different calibrated spectra than the stisextract method \citep{how00} we used
in previous krypton analyses \citep{car01,car03}, a comparison of the
abundances derived when the data are calibrated using each method indicates
that the results generally agree within 0.02 dex. In view of this accord, and
since it is more efficient to process the data through a single algorithm, we
calibrated the new data using CALSTIS alone. 

In order to alleviate concerns of unresolved saturation, we followed the same
abundance determination procedures used in our previous work. Specifically,
both apparent optical depth (AOD) and profile-fitting methods were applied to
the data to check consistency, then the profile-fitting results were adopted
while AOD uncertainties were folded into the listed column density errors to
explicitly account for continuum placement uncertainty. For the majority of
the new sight lines, the spectral coverage included wavelengths from shortward
of 1200 {\AA} to longward of 1250 {\AA} so that the neutral hydrogen column
density could be assessed by continuum reconstruction near the Ly-$\alpha$
line. {\it FUSE} data were also available for nearly all of these paths, so
the same method was used to determine the molecular hydrogen column density
toward each target star. The krypton and hydrogen abundance data for the newly
identified krypton sight lines are given in Table~\ref{KrHdata}. For more
detail on the techniques mentioned here in brief, please see \citet{car03} and
references therein. All previously-published krypton and hydrogen abundances
for sight lines where \ion{Kr}{1} $\lambda$1236 has been detected are compiled
in Table 2 of that paper.

\section{The Gas-Phase Kr/H Ratio Distribution}
\label{krypton_distribution}
The mean interstellar krypton gas-phase abundance level that \citet{car97}
derived, represented by the Kr/H ratio log$_{10}$(Kr/H) = $-9.02\pm0.02$ [Kr/H
= $(0.96\pm0.05)\times10^{-9}$], was remarkable in that each and every
measurement of the ratio agreed with this value within its 1$\sigma$ error bar.
Nine of the ten spectra compiled in that study were observations of stars
within 500 pc of the Sun, indicating that the ISM within this region is very
well mixed. This evidence for an homogeneous local ISM was supported by the
analysis of GHRS observations of two \ion{N}{1} absorption lines near 1160 {\AA}
and an \ion{O}{1} line at 1356 {\AA} in the spectra of stars with similar
distances, many of them the same stars toward which krypton was detected.
Gas-phase abundance levels of log$_{10}$(N/H) = $-4.12\pm0.02$ \citep{mey97}
and log$_{10}$(O/H) $-3.46\pm0.02$ (\citealt{mey98}; this value has been
adjusted to reflect updates in the oxygen line's oscillator strength) were
identified by these studies and although both nitrogen and oxygen, unlike
krypton, are susceptible to depletion from the gas phase, no deviation from the
interstellar means by more than 1.5$\sigma$ were evident. These data indicated
that the degree of mixing in the local ISM is independent of the elements'
different nucleosynthetic origins.

\citet{car01} added 10 STIS-observed spectra into the discussion of krypton
abundance, probing denser and longer sight lines. Regardless of the much
broader volume of parameter space being studied, in terms of total column
density, ${\langle}n_{\rm H}\rangle$, extinction, and distance, each of their
new Kr/H data points also agreed with the unchanged overall interstellar mean
within 1$\sigma$. Most recently, \citet{car03} added 12 more paths to the
sample. Their data indicated that among the 22 sight lines less than $\sim$800
pc in length, all but one were consistent with a singular interstellar Kr/H
ratio, while the exception deviated by no more than 1.3$\sigma$. Of the 4 paths
between 1.5 and 4.8 kpc in length, however, only the two directed along the
curve of the Orion Spur (the local spiral arm situated between the
Carina/Sagittarius and Perseus Arms) agreed with the otherwise ubiquitous Kr/H
ratio already established. The exceptional sight lines, toward HD 116852 and
HD 152590, both intersected the Carina/Sagittarius Arm and it was suggested that
their roughly solar krypton abundances might be related to this property. It
should be noted, however, that the abundance measurements for these paths were
among the most uncertain, given complications with distinguishing weak
absorption components from the noise in the HD 116852 spectrum and the unusually
steep curvature of the HD 152590 stellar background near 1236 {\AA}.

\begin{figure}
\epsscale{1.0}
\plotone{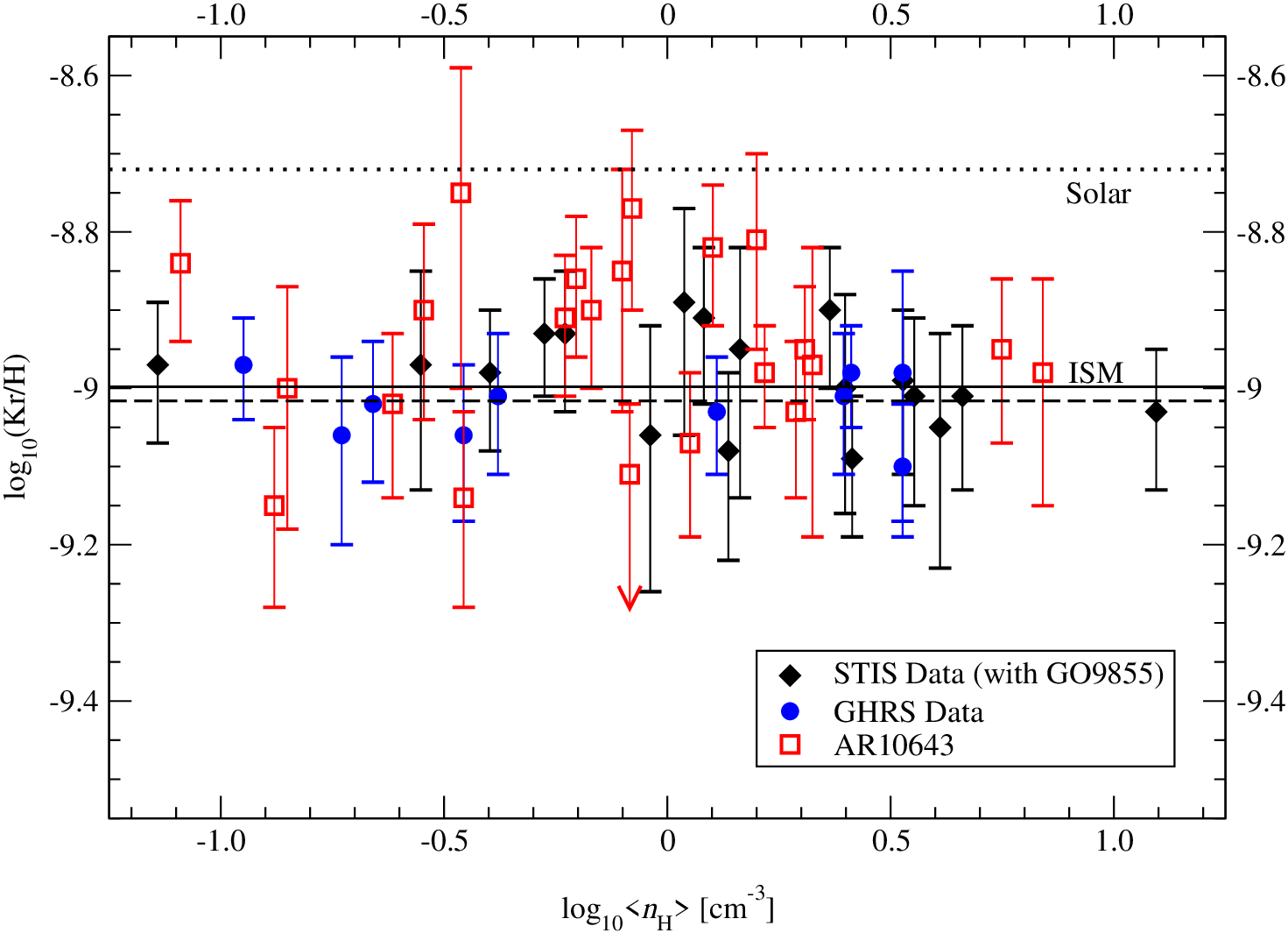}
\caption{Density-dependent gas-phase interstellar krypton abundances. New
krypton-to-hydrogen abundance ratios for 26 sight lines overlap
previously-published GHRS and STIS measurements for 22 other paths, when
expressed as a function of mean hydrogen sight line density. The new data are
scattered more widely with respect to the ISM weighted mean of
log$_{10}$(Kr/H) = $-9.00\pm0.02$ than were the previous data to their mean;
the two weighted averages agree within their respective errors. The new and old
[log$_{10}$(Kr/H) = $-9.02\pm0.02$; \citealt{car03}] interstellar Kr/H
abundance ratios are indicated by the solid and dashed lines, respectively; the
dotted line is the \citet{lod03} Solar value [log$_{10}$(Kr/H) =
$-8.72\pm0.08$].
\label{krnhplot}
}
\end{figure}

Both stars were re-observed for GO9855, with the following results:
log$_{10}$(Kr/H)$_{\rm HD 116852}$ = $-8.97^{+0.08}_{-0.09}$;
log$_{10}$(Kr/H)$_{\rm HD 152590}$ = $-8.93^{+0.06}_{-0.07}$. The new
observations have provided spectra integrated for more than 15 and 7 times the
previous exposure times, respectively. As a result, the difficulties with the
first measurements of krypton abundance along these sight lines were
dramatically reduced. Perhaps unsurprisingly, the Kr/H ratios for both sight
lines are closer to the interstellar mean than previously determined; the
krypton abundance toward HD 116852 matches log$_{10}$(Kr/H)$_{ISM}$ =
$-9.02\pm0.02$ within error, while the HD 152590 path is only the second, after
HD 37367, to deviate by as much as an amount approaching 1.3$\sigma$. The
remaining two sight lines successfully observed for the GO9855 program, HD
40893 and HD 165246, agree within their respective uncertainties with the
interstellar mean krypton abundance (see Table~\ref{KrHdata}). So before the
results from the STScI Archive (by way of AR10643) are considered, the picture
we have constructed is of a remarkably uniform sample of 28 Kr/H ratios which
differ from a single value by no more than 1.3$\sigma$ (see
Figure~\ref{krnhplot}). Since krypton is undepleted along the sight lines that
can be probed using ultraviolet absorption features, this tight distribution is
strong evidence that the diffuse ISM is very well mixed on length scales of
several hundred parsecs. Combining the krypton result with data for other
elements (e.g., O, Mg, P, Mn, Ni, Cu, Ge) has implied that the composition of
the diffuse ISM is homogeneous to a limit of order 0.04 dex \citep{car06} on
this same spatial scale.

\begin{figure}
\epsscale{0.9}
\plotone{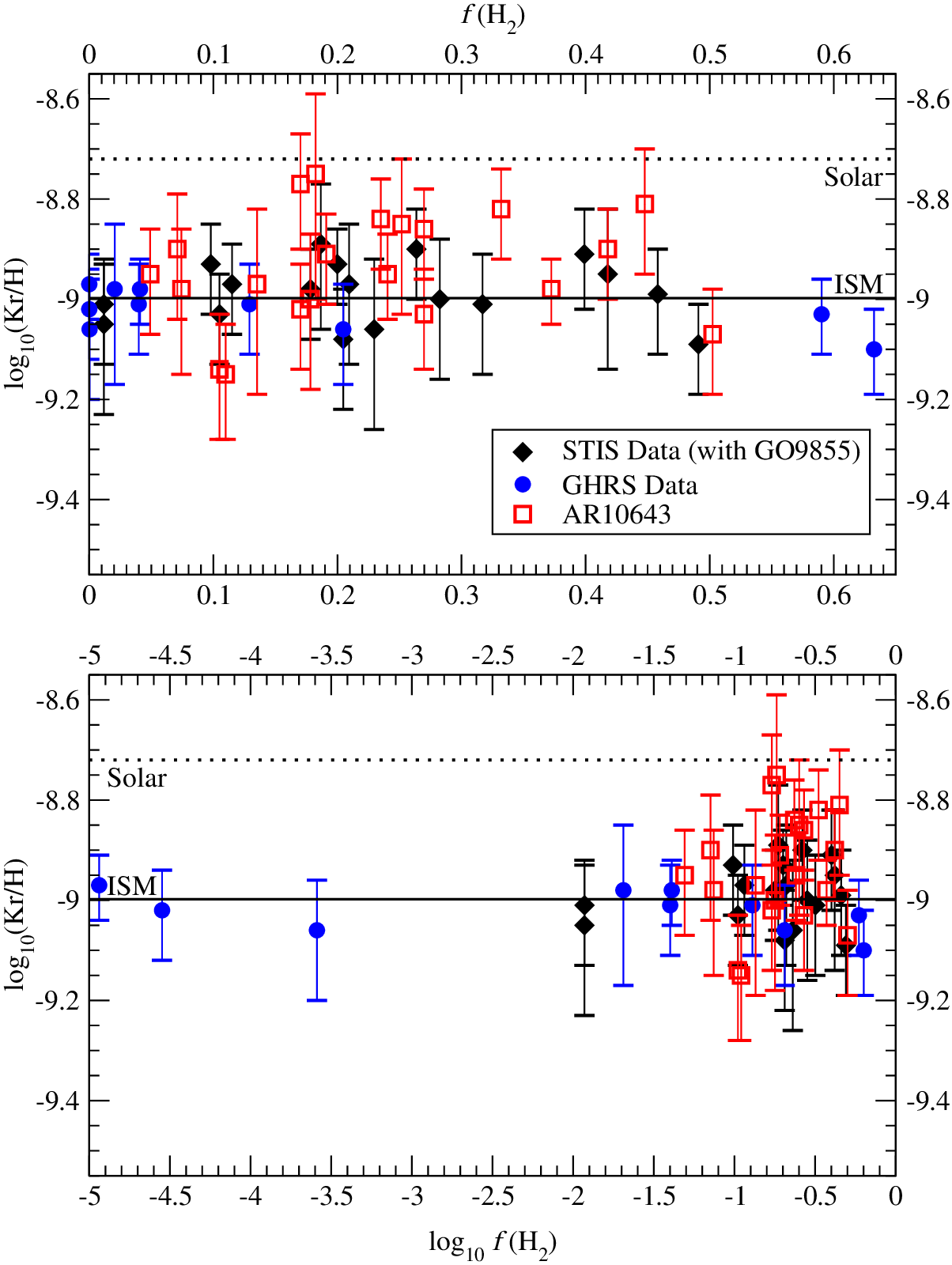}
\caption{Krypton abundance and molecular hydrogen fraction. The gas-phase Kr/H
abundance ratio is shown as a function of molecular hydrogen fraction in the
top panel and its base-ten logarithm in the bottom panel. It is clear in the
top panel that from low to high values of $f$(H$_2$), the data are consistent
with a uniform Kr/H ratio; the bottom panel emphasizes that this uniformity
extends across 5 orders of magnitude to very low values of $f$(H$_2$). These
plots confirm the hypothesis that krypton is undepleted in the diffuse ISM.
\label{krh2plot}
}
\end{figure}

\begin{figure}
\epsscale{1.0}
\plotone{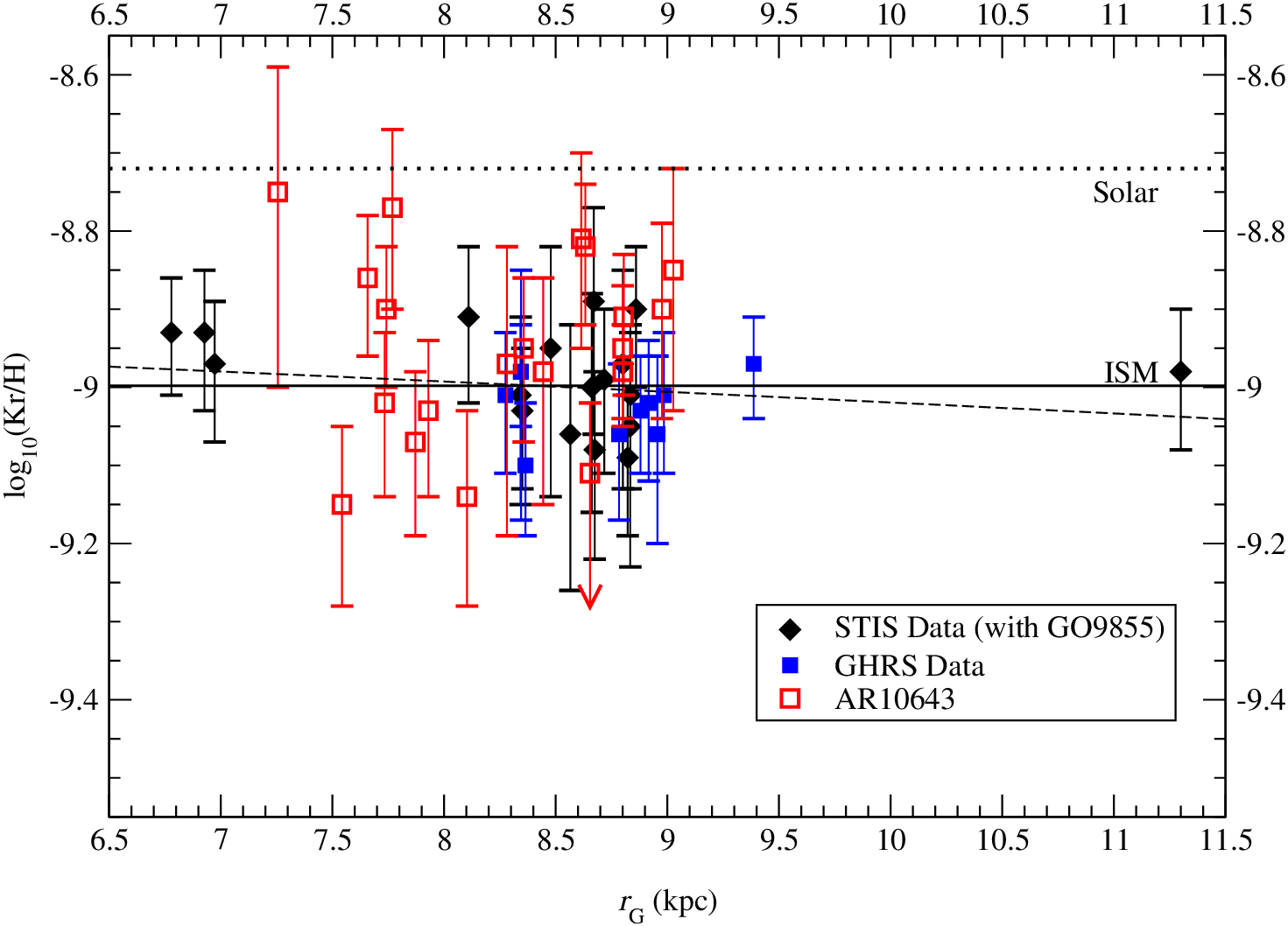}
\caption{Krypton abundance and Galactocentric radius. The gas-phase Kr/H
abundance ratio does not exhibit a strong variation as a function of the target
star's distance from the Galactic centre; the mean slope is $-0.013\pm0.014$
dex/kpc when a linear function is fit to the data (the dashed line). This value
suggests a radial abundance gradient of order $-$0.027 dex/kpc under the crude
assumption that on average the gas probed in each sight line is situated midway
along its pathlength. This trend is not very visually compelling, however,
particularly since the datapoints within 8.5 kpc of the Galactic center that
have the two largest Kr/H values are subject to the noted heliocentric distance
trend.
\label{krrgplot}
}
\end{figure}

\begin{figure}
\epsscale{1.0}
\plotone{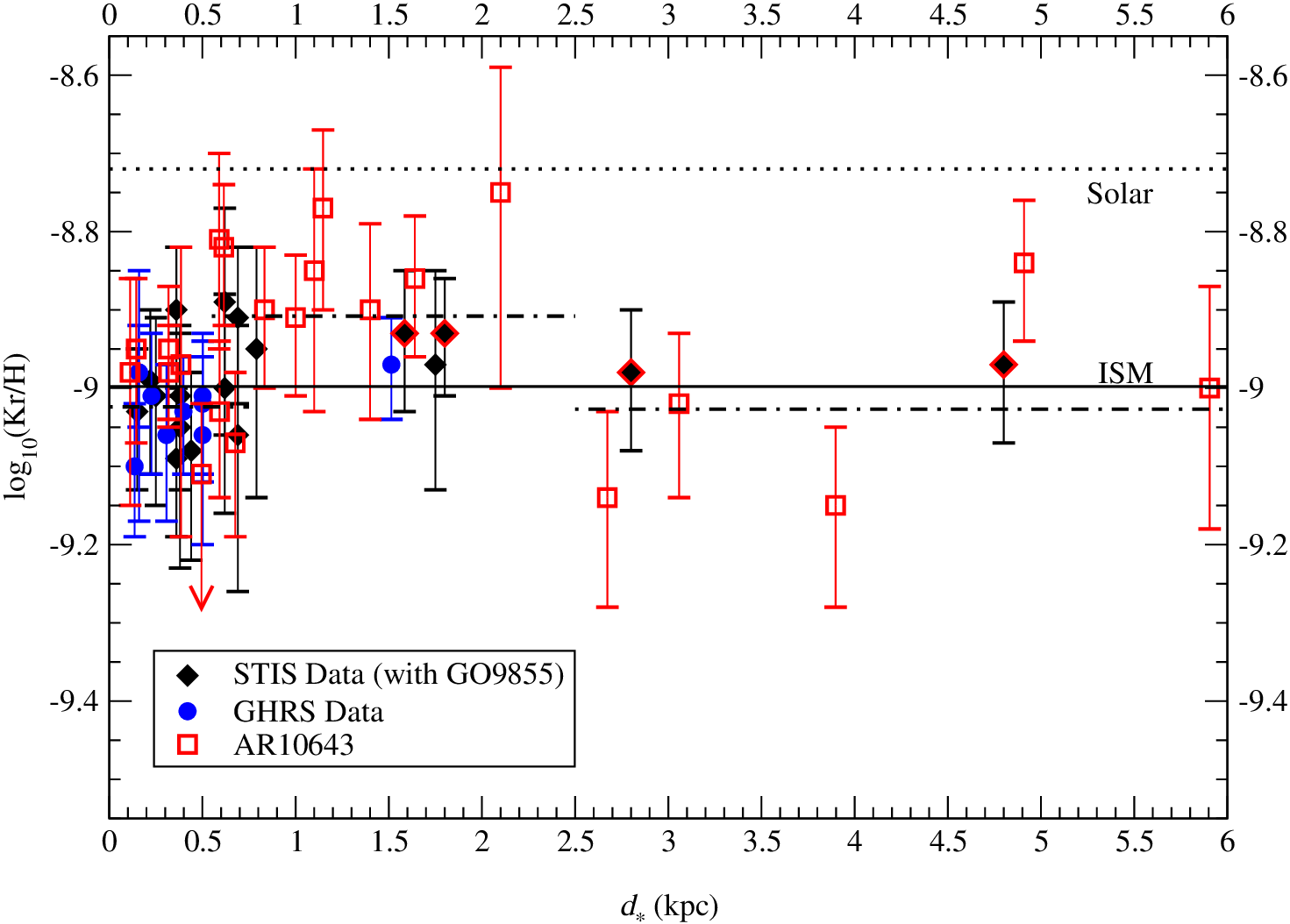}
\caption{Distance-dependent gas-phase interstellar krypton abundances. The
krypton-to-hydrogen abundance ratios for all 48 sight lines are plotted as a
function of the distance to the target star. The new data include many distant
paths, separating the sample into paths with low Kr/H ratios within 750 pc
[weighted mean log$_{10}$(Kr/H) = $-9.02\pm0.02$], those with high Kr/H ratios
and lengths between 550 and 2500 pc [weighted mean log$_{10}$(Kr/H) =
$-8.91\pm0.02$], and several longer than 2.5 kpc whose Kr/H ratios return to
values consistent with those in the local ISM [weighted mean log$_{10}$(Kr/H) =
$-9.03\pm0.04$]. The weighted mean Kr/H value in each circular region are
indicated by the dot-dashed lines.
\label{krdpplot}
}
\end{figure}

However, this krypton abundance sample of 28 sight lines is dominated by short
paths extending less than a kiloparsec through the Galactic disk. To more fully
comment on the singular krypton level, and hopefully identify regions with
distinct nucleosynthetic histories, it became necessary to consider all
possible unpublished \ion{Kr}{1} $\lambda$1236 detections in the Archive. The
data from Table~\ref{KrHdata} and all previous krypton abundance measurements
have been encapsulated in Figures~\ref{krnhplot}, \ref{krh2plot},
\ref{krrgplot} and \ref{krdpplot}, which describe the fluctuations in Kr/H with
the sight line properties mean sight line hydrogen density [${\langle}n_{\rm
H}\rangle$], molecular hydrogen fraction [$f$(H$_2$)], Galactocentric radius
($r_{\rm G}$), and sight line pathlength ($d_\ast$), respectively.

The first two plots serve to re-iterate and emphasize previous conclusions
regarding the state of krypton in the interstellar medium. In
Figure~\ref{krnhplot}, the Kr/H gas-phase abundance ratio distribution does not
depend on ${\langle}n_{\rm H}\rangle$, the mean hydrogen sight line density.
Previously-published data are more uniform than the newer data in their
collective agreement, differing by less than 1.3$\sigma$ from the mean on a
point-by-point basis, but the 26 Archive-based Kr/H ratios were derived from
spectra not designed to measure krypton abundance and so generally suffer from
larger uncertainties and are more widely distributed. Nevertheless, upon
including these new data the overall interstellar mean shifts only slightly to
log$_{10}$(Kr/H)$_{ISM}$ = $-9.00\pm0.02$ [or Kr/H =
$(1.00\pm0.04)\times10^{-9}$] from the previous determination, and there are no
coherent trends with ${\langle}n_{\rm H}\rangle$. Figure~\ref{krh2plot} clearly
shows that the measured Kr/H ratio is also independent of a sight line's
molecular hydrogen fraction. This property is evident whether the molecular
hydrogen fraction is expressed strictly as a number, emphasizing the uniform
distribution of Kr/H ratios for $f$(H$_2$) $> 0.1$, or in logarithmic form,
which stresses the singular nature of the Kr/H ratio across several orders of
magnitude to very small values of $f$(H$_2$). In other words, the shielded
conditions that at the same time favor the enhancement of H$_2$ abundances
relative to \ion{H}{1} and elemental depletion into dust grains do not affect
the gas-phase abundance of krypton. Thus krypton is undepleted in the ISM,
within the density range and spatial regions probed to date by Galactic sight
lines.

The broad density-parameter strokes used to define this portrait of
kiloparsec-scale krypton abundance uniformity and lack of depletion into dust,
however, mask some subtler features. In particular, the scatter among the older
measurements is only 0.06 dex, compared with a typical Kr/H ratio uncertainty
of 0.09 dex. Encompassing the new and Archival data into the sample increases
this scatter to 0.10 dex. And although the Archival Kr/H ratios are in general
less precise than earlier measurements, it should be noted that they are also
largely derived from longer paths through the Galactic disk. In fact, according
to Figures~\ref{krrgplot} and \ref{krdpplot}, the increased Kr/H ratio scatter
is not due solely to greater inherent uncertainty in the Archival measurements,
but is perhaps dominantly produced by variations dependent on spatial sight
line properties.

Consider Figure~\ref{krrgplot}, which examines Kr/H as a function of the
galactocentric radius ($r_G$) of the observed OB star. In studies of stellar
atmospheres (e.g., \citealt{rol00}) and planetary nebulae (e.g.,
\citealt{mar02}), a clear abundance trend with $r_G$ is apparent. As a star's,
or planetary nebula's, distance from the Galactic center increases, elemental
abundances steadily fall. For example, the oxygen radial abundance gradient is
determined by these studies to be about $-0.07$ dex/kpc, attributed to enhanced
nucleosythesis in the central regions of the Galaxy relative to outlying areas.
It is more difficult to measure such trends using abundances determined from UV
absorption line profiles because these profiles are convolutions of features
from individual clouds along the line of sight to which the distances are
rarely known. Nevertheless, the Kr/H data are plotted as a function of $r_G$ in
Figure~\ref{krrgplot} to search for this trend. Unfortunately, only a few stars
are situated beyond 500 pc from the Sun's radial location in the disk, either
closer to or farther from the Galactic center (the adopted Solar galactocentric
radius for this analysis is 8.5 kpc). Moreover, the bulk of stars outside the
annulus are to be found nearer the Galactic centre than away from it. A
weighted linear regression to the radial dependence of Kr/H gives a slope of
$-0.013\pm0.014$ dex/kpc, which becomes a krypton radial abundance gradient of
$-$0.027 dex/kpc under the crude approximation that all absorbing material is
situated at the midpoint of each sight line (i.e., the radial gradient is
double the measured slope). The slope, however, includes zero within its
1$\sigma$ uncertainty and the trend is not very visually compelling.
Consequently, these data are also consistent with a flat gradient.

The small apparent slope with $r_G$ may simply be a secondary result of the
striking spatial pattern in krypton abundance that becomes clear when Kr/H is
plotted against the distance to the observed star, as in Figure~\ref{krdpplot}.
Previously published data are consistent with krypton abundance being
independent of the target distance, yet they conform to a pattern of elevated
Kr/H ratios for 750 pc $< d_* <$ 2500 pc when combined with measurements of
features found in Archival spectra. Note that the four GO9855 sight lines are
those black diamonds limned in red and that they are consistent with both
patterns. There is some overlap for targets at distances between 550 and 750 pc
from the Sun, where a fraction of the sight lines exhibit elevated krypton
abundances and the remainder match the level of the previously published
interstellar mean. The dot-dashed lines in this plot indicate the mean Kr/H
ratios for each distance interval. If the sight lines with pathlengths between
550 and 750 pc are allotted by their krypton abundances to the groups with
pathlengths less than 550 pc and those between 750 and 2500 pc long, then paths
with endpoints in the defined annulus are elevated in Kr/H by 0.11 dex with
respect to the local ISM mean (see Table~\ref{aratmeans}). Notably, the Kr/H
ratios interior and exterior to the annulus are identical within their errors,
each differing from the value within the annulus by roughly 3$\sigma$.

\begin{figure}
\epsscale{1.0}
\plotone{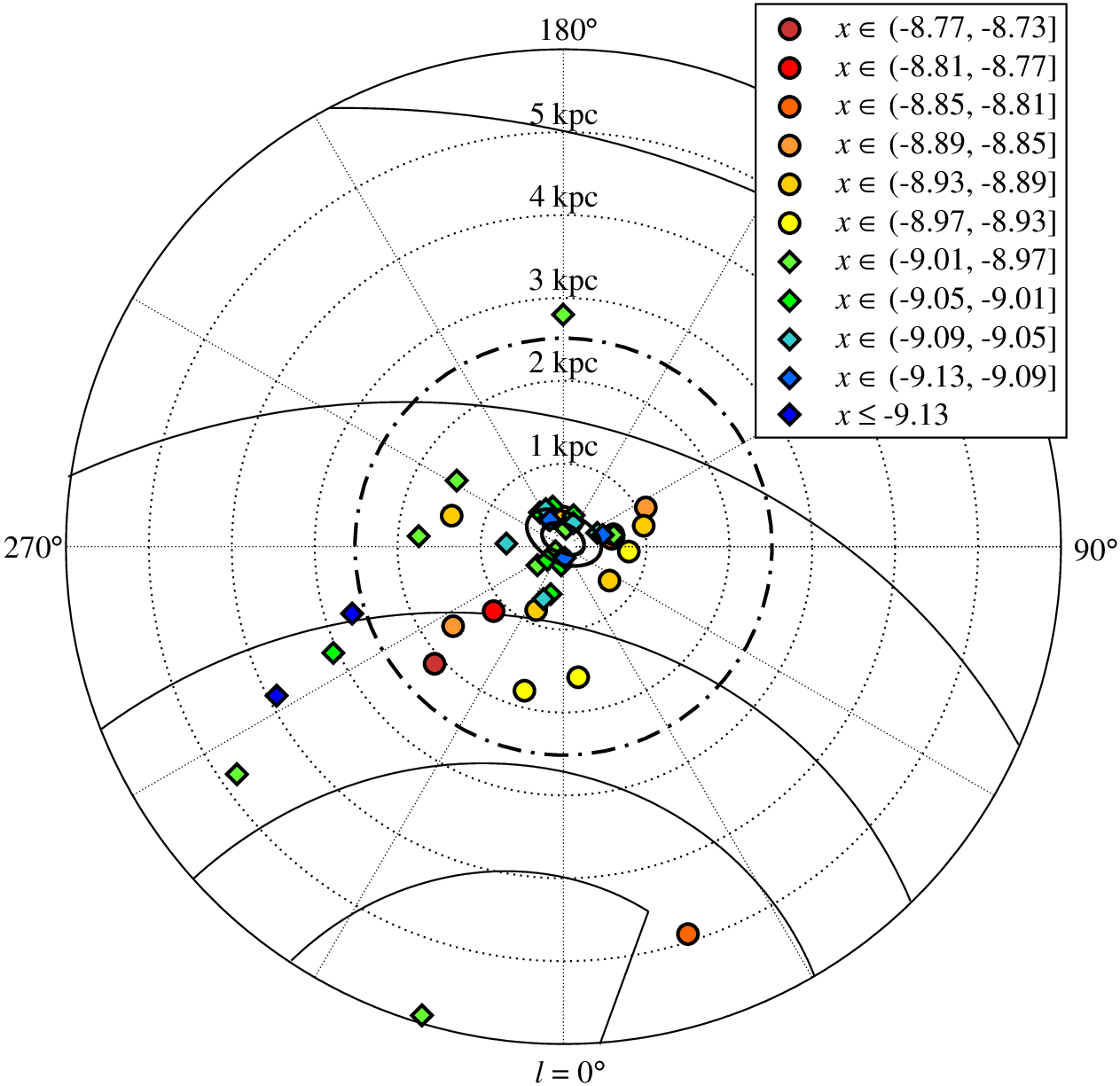}
\caption{The Galactic plane distribution of gas-phase interstellar krypton
gas-phase abundances. The krypton sight line endpoints are identified by
symbols indicating the corresponding Kr/H abundance ratio relative to the
interstellar mean. The solid curves indicate spiral arm solutions \citep{val02}
and the dotted annuli specify heliocentric distance in kiloparsecs. In this
plot, the area near the Sun is dominated by low abundance ratios, surrounded by
a rough annulus of higher than average ratios. On average, sight lines longer
than 2.5 kpc return to lower abundances. The dot-dashed circle marks the outer
edge of the annulus of elevated Kr/H ratios we have noted; the solid ovals near
the center of the plot indicate the area bounded by projection of the Gould
Belt onto the Galactic plane, roughly identifying both the inner and outer
dimensions \citep{per03}.  In the figure legend, $x$ refers to log$_{10}$(Kr/H).
\label{krradplot}
}
\end{figure}

Although the difference between weighted Kr/H ratio means for the annulus and
non-annulus sets provide strong evidence that the Kr/H abundance ratio changes
with heliocentric distance, the effect is just at the 3$\sigma$ level and the
sample sizes are small. To quantitatively confirm the reality of the annulus of
elevated krypton abundances, we have applied the Student's t-test to the data.
The t-test is a statistical measurement that determines the likelihood that two
datasets are merely random samplings of the same underlying population---in
this case that of Kr/H abundance ratios. Expressing the data as sets of
log$_{10}$(Kr/H) values, we find that the annulus and non-annulus groups are
both reasonable approximations of normal distributions, that the probability
that they are derived from the same underlying population is 0.000034\%, and
that at the 95\% confidence level the difference between the means for these
two samples is at least 0.09 dex. Therefore, we conclude that these groups of
sight lines indeed represent distinct populations. This analysis blends
seamlessly with a more qualitative description of the distinctions between the
two distributions. Specifically, the breadth of the krypton abundance
distribution interior to the annulus is characterized by a scatter of 0.06 dex,
indicating a particularly tight grouping given the typical uncertainty of 0.09
dex in this measurement. The scatters for the groups of sight lines both ending
within the annulus and exterior to it are larger, reflecting the larger typical
uncertainties for the corresponding paths, but the annulus group is still
scattered to a lesser degree than would be indicated by the typical abundance
ratio uncertainty. Thus, an annulus of inner radius roughly 600 pc centered
near the Sun defines a region of enhanced krypton abundance relative to the
interstellar mean for the Galactic disk within 6 kpc. See
Figure~\ref{krradplot} for an illustration of this region in terms of Kr/H
abundance ratios, looking down on the Galactic plane from above.

\section{Is the Effect Only Evident for Krypton?}
\label{distance_effect}
This annulus of krypton abundance enhancement has previously been detected in
the abundance data for other elements, although its interpretation in these
instances has been different. Two independent compilations of O/H abundance
ratios based on the weak $\lambda$1356 intersystem line \citep{and03,car04}
noted that this quantity is generally constant except for two effects: enhanced
depletion along dense sight lines and an elevation in the ratio for paths
longer than about 800 pc and less dense than ${\langle}n_{\rm H}\rangle = 1.0
{\rm cm}^{-3}$. \citet{car06} showed that copper to hydrogen abundance ratios
also exhibit the latter dependence on a sight line's length. Unfortunately, the
database of element-to-hydrogen abundance ratios that they compiled for the
other elements being studied (Mg, P, Mn, Ni, and Ge) did not include enough
short low-density sight lines to rigorously establish that these elements also
conformed to the same trend, and including denser sight lines would have
confused any distance effect with the observed depletion enhancement with
density. Nevertheless, there was some support for elevated abundance ratios
beyond 800 pc in these data as well.

The interpretation of the heliocentric distance trend for these studies and
the earlier GHRS investigations has been guided by simple spatial arguments.
When the first high spectral resolution data from GHRS were compiled, the goal
was to explain the apparent constancy of interstellar C, O, N, and Kr
abundance levels within 500 pc of the Sun at a common level relative to their
Solar values. \citet{mey98} posited three possible explanations: early
solar system enrichment by a local supernova; a recent infall of metal-poor gas
in the local Milky Way; or an outward diffusion of the Sun from a smaller
Galactocentric distance. Each possibility would have the effect of reducing
elemental abundances in the Solar neighborhood by a common factor relative to
the Sun, provided that radial abundance gradients for these elements were also
similar. For the effects observed by both oxygen compilations and
\citet{car06}, the abundance reduction is not tied to the Sun, so the simplest
mechanism to reduce abundances by a common factor over a large region, and
moreover reduce the abundances of several elements by the same factor, is
recent infall of metal-poor gas.

For other abundance analyses, however, alternate explanations have also been
considered. The abundance of nitrogen, ranking along with oxygen as one of the
five most abundant elements in the galaxy, also exhibits unexpected variations
in the local ISM, for which possible explanations may be nucleosynthetic or
related to differential mixing. Specifically, GHRS observations of the weak
\ion{N}{1} 1160 {\AA} doublet described a constant N/H ratio within 500 pc of
the Sun, with negligible depletion into dust grains \citep{mey97}, but {\it
Interstellar Medium Absorption Profile Spectrograph}, STIS, and {\it FUSE} data
have altered this picture by enlarging the sample to probe more diverse
environments. From these data, \citet{kna03} found that N/H varied
significantly as a function of $N$(H) without showing any evidence of N$_2$
formation or depletion into dust. Combining nitrogen and oxygen abundances for
the most robustly determined of these N column densities, \citet{kna06}
determined that locally the N/O ratio is 0.217$\pm$0.011, but that beyond 500
pc the ratio drops to 0.142$\pm$0.008 . Eight of the 13 sight lines from which
this pattern is deduced are also among those with krypton abundances, and the
two paths added in proof to that paper bring the total to ten whose Kr/H ratios
can be checked against the distance pattern noted here. Toward each star that
was reported by \citet{kna06} as being situated within 500 pc of the Sun, the
krypton abundance is in agreement with the local mean [log$_{10}$(Kr/H)$_{ISM}$
= $-9.02\pm0.02$]. And furthermore, those paths assigned lengths between 750
and 2500 pc agree with log$_{10}$(Kr/H)$_{ISM}$ = $-8.91\pm0.02$, the mean
abundance for the annulus of enhancement. In this paper, however, we have
adopted a distance of 3.06 kpc for HD 99857 rather than the sub-500 pc {\it
Hipparcos} value used by \citet{kna06}. The parallax error on the latter value
is of order 80\%, and the distance we adopt is more consistent with the star's
spectroscopic parallax and previous literature estimates of its distance. But
regardless of whether the path is less than 500 pc or longer than 2.5 kpc, the
krypton abundance means in these two regions are in agreement (see
Table~\ref{aratmeans}).

A similar distinction between local ISM abundances and those determined using
paths longer than a few hundred pc has been noted for deuterium, although the
transition distance cannot be as cleanly defined. \citet{woo04} compiled D/H
measurements from a variety of instruments, finding that both within 100 pc of
the Sun and beyond about 600 pc the ratio was well-defined but that the levels
were roughly a factor of two (0.30 dex) different. Between 100 and 600 pc,
there is a complicated distribution including data spanning a much wider range
than is delineated by the near and far average D/H values. Subsequent analyses
have expanded upon this description with new data, considering scenarios of
infall, variable astration, and density-dependent depletion as explanations for
the significant D/H ratio variability among paths between 100 and 600 pc long
(e.g., \citealt{oli06,lin06}). However, the possibility of infall has been
regarded as doubtful for that situation because it would need to be confined to
particular individual sight lines, both since the theoretical ISM mixing
timescale of 350 Myr suggests that variabilities are likely to have been
removed on these lengthscales in the same time as those within 100 pc of the Sun
\citep{dea02} and because empirical evidence suggests that mixing is efficient
in this region of the ISM \citep{car06}. Nevertheless, it is apparent that the
interstellar deuterium abundance also is subject to anomalies dependent on
heliocentric distance, and that the scale of the pattern is similar to that
evident in oxygen, copper, and other elements, particularly since the lengthy
Galactic sight lines of D/H studies generally would meet the definition of the
annulus group in this paper.

The spatial extent of the abundance enhancement in heavy elements is made
obvious with the current krypton data because measurement of the gas-phase
abundance of this element is not subject to complication by depletion---UV
absorption profiles permit the {\it total} amount of neutral krypton along a
given sight line to be determined. For elements that play a role in
interstellar dust, spatial variations in abundance can be rivalled or exceeded
in magnitude by those related to density-dependent depletion. Using oxygen as
an example, the distinction between the derived warm and cold ISM depletion
levels is 0.10 dex, comparable to the location-based abundance enhancement
identified here in krypton. This complication made it easy to miss the trend in
the \citet{car04,car06} oxygen and copper data; specifically, to note that
beyond a distance of order 2.5 kpc, elemental abundances largely return to the
lower levels measured within 750 pc of the Sun. In order to confirm the reality
of this spatial abundance pattern, though, it is necessary to show that the
effect is not just distinguishable in the separate sight line samples for more
than one element, but that it is apparent using the same sight lines for each
element being considered.

\begin{figure}
\epsscale{0.7}
\plotone{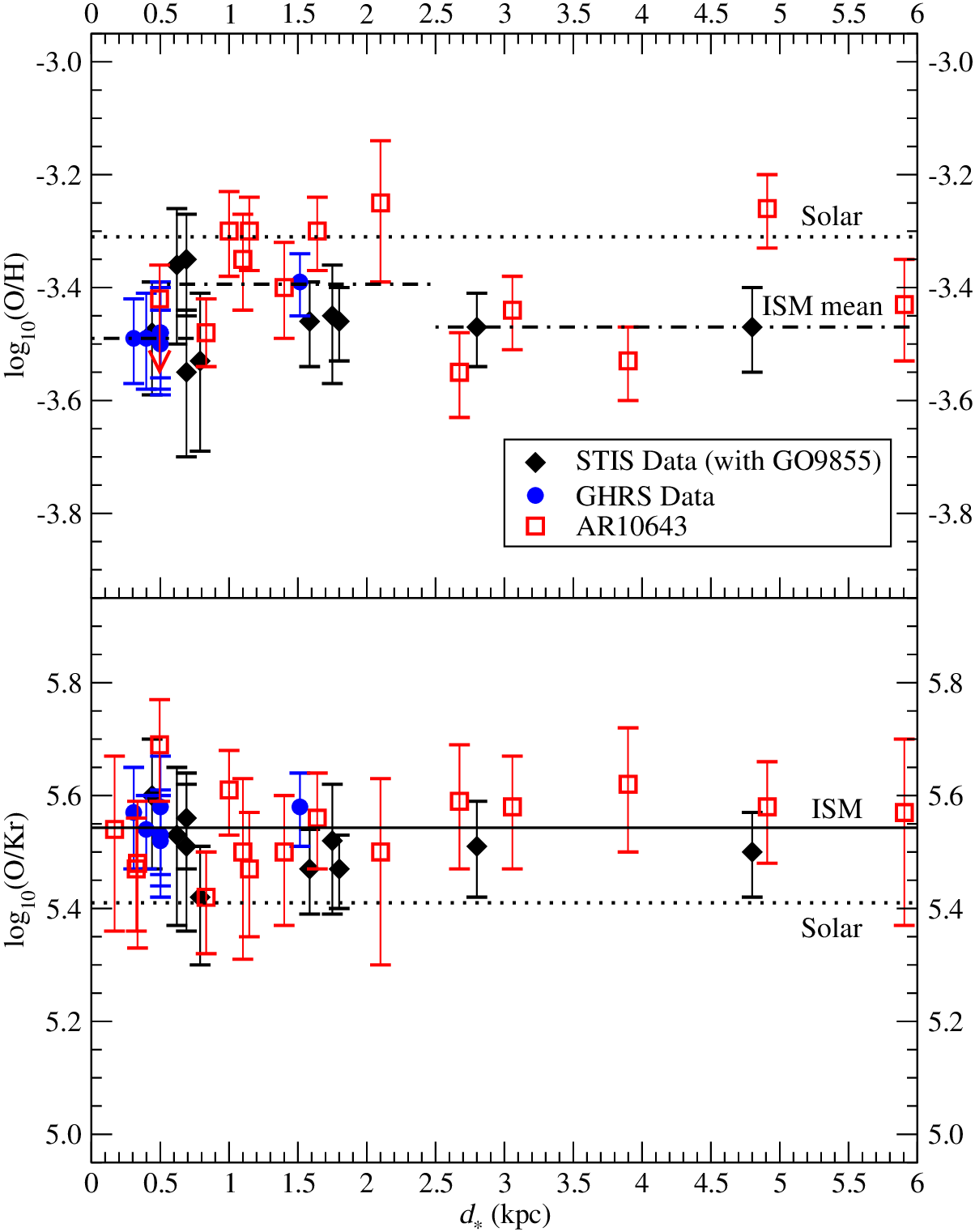}
\caption{Distance-dependent gas-phase interstellar oxygen abundances along
paths with ${\langle}n_{\rm H}\rangle < 1.0 {\rm cm}^{-3}$. Gas-phase oxygen
abundances for the current krypton sample are presented above as a function of
sight line pathlength. In the top panel, the data are displayed as O/H ratios;
an enhancement is clearly evident for paths between 750 and 2500 pc in length,
mirroring the trend in Kr/H abundance ratios from Figure~\ref{krdpplot}. This
trend is approximately equal in strength for krypton and oxygen, as indicated
by the extremely tight distribution of O/Kr abundance ratios in the bottom
panel. Note that only low-density sight lines [${\langle}n_{\rm H}\rangle
< 1.0 {\rm cm}^{-3}$] are used for these plots, to avoid confusing the clarity
of the effect by including paths affected by oxygen depletion \citep{car04}.
Numerical values for the ISM O/H means are found in Table~\ref{aratmeans}; the
Solar ratios are from \citet{lod03}.
\label{okrdpplot}
}
\end{figure}

Since the distance effect has previously been observed in gas-phase O/H
abundance ratios and \ion{O}{1} $\lambda$1356 features are similar in optical
depth to \ion{Kr}{1} $\lambda$1236, given their respective interstellar
abundances, we measured the oxygen abundance for each sight line presented in
this paper. The oxygen measurements were made using the same techniques, with
the same underlying profile solutions, applied to the corresponding krypton
features and the results are presented in Table~\ref{OHdata}. Note that in
cases where the profile is very broad, oxygen abundances are only tallied over
radial velocities consistent with the krypton profile, so that O/Kr ratios are
representative of the reliable features in each spectrum. Also, the ionization
potential of neutral oxygen (13.618 eV) is very close to that of krypton and
just larger than the corresponding hydrogen value, ensuring that oxygen and
krypton will both be neutral in neutral gas. Care has been taken when plotting
the gas-phase O/H ratios in the top panel of Figure~\ref{okrdpplot} to avoid
paths with ${\langle}n_{\rm H}\rangle > 1.00 {\rm cm}^{-3}$, above which the
data are likely to be contaminated by enhanced oxygen depletion into grains
\citep{car04}. Using the same groupings adopted in the analysis of Kr/H
variation with pathlength, O/H and O/Kr abundance ratio means for each distance
range are given in Table~\ref{aratmeans}; the individual O/Kr abundance ratios
are plotted in the bottom panel of Figure~\ref{okrdpplot}. Notably, the same
distance pattern identified in Kr/H can easily be discerned in gas-phase O/H
ratios using the same sight lines. Numerically speaking, the two abundance
ratio increases are also similar: the Kr/H ratio is elevated by 0.11 dex
($\sim4\sigma$, using the quadrature sum of the uncertainties in annulus and
non-annulus weighted means) for sight lines ending within the 550---2500 pc
annulus relative to the value both inside and outside, while the gas-phase O/H
ratio is elevated by 0.08 dex (2.5$\sigma$) in the same manner.

Underlining this evidence for a nearly identical heliocentric distance
dependence for the krypton and oxygen abundances relative to hydrogen is an
extremely tight distribution when the ratio of oxygen to krypton is plotted
against pathlength in the bottom panel of Figure~\ref{okrdpplot}. Leaving out
sight lines subject to enhanced oxygen depletion, but including those paths
with significant krypton detections excluded from the Kr/H analysis for lack of
rigorously-determined hydrogen column densities (e.g., HD 94454), the overall
weighted mean O/Kr is $5.55\pm0.02$ with a scatter of 0.06 dex. It should be
noted that the breadth of this distribution is significantly smaller than what
would be expected given the typical abundance ratio uncertainty for these data
($\sim$ 0.09 dex). Similar numbers were derived by \citet{car03} for the Kr/H
ratio, and used by \citet{car06} to demonstrate that the ISM within several
hundred parsecs of the Sun is very well mixed. Delving further into the details
of Table~\ref{aratmeans}, the mean gas-phase O/Kr abundance ratios for paths
both entirely within the annulus and those extending well beyond it agree
within their errors at a level somewhat larger than the mean for paths in the
annulus sample; however, the annulus and non-annulus values agree within the
quadrature sum of their 1$\sigma$ errors. Consequently, the abundance
enhancement observed in krypton for sight lines between 750 and 2500 pc in
length is also proven detectable in oxygen, and at a similar level.

Combining these results, it appears that the spatial abundance enhancement is
nucleosynthetic in nature. While krypton, oxygen, and copper abundances all
rise by a similar amount among paths with a characteristic range of lengths,
nitrogen apparently does not. Specifically, we infer that if the O/H ratio is
elevated for the annulus sample by 0.08 dex, the N/H ratio for the same sight
lines must {\it fall} by 0.10 dex in order to reproduce the mean N/O abundance
ratios observed by \citet{kna06}. The scenario suggested by these properties
would be described as follows: paths shorter than 550 pc would probe a
well-mixed ISM; local ISM gas intermingles with material that has experienced
more recent nucleosynthetic activity at distances from the Sun beyond about
500 pc, to the point where this more heavy-element-rich gas dominates in a ring
perhaps a few hundred parsecs wide; finally, more than 1000 pc away from the
Sun, the ISM appears to again resemble the local ISM in terms of elemental
abundance.

Such a configuration would give rise to the observed abundance ratio
dependencies, given that the total hydrogen column densities for the entire
database of sight lines with krypton detections range only from log$_{10}N$(H)
= 20.46---21.77 and about 80\% of these are within a factor of 4 of one
another, ranging from log$_{10}N$(H) = 21.20---21.77 . Specifically, paths
shorter than 550 pc and some with lengths in the 550---750 pc range would be
dominated by the well-mixed local ISM. Other sight lines up to 750 pc long
would intersect gas dominated by more recently processed material. The paths in
this dataset with lengths of 750---2500 pc are exclusively of lower densities
(${\langle}n_{\rm H}\rangle < 1 {\rm cm}^{-3}$) and their elemental abundance
ratios might readily be dominated by processed gas concentrated in an annulus
centered near the Sun whose inner radius is about 500---600 pc. And beyond 2500
pc, sight lines with similar overall $N$(H) to shorter paths are on average
more rarified and less likely to be dominated by gas in the implied ring, and
so evince elemental abundance ratios apparently characteristic of both the
well-mixed local gas and more distantly situated material.

The suggested annulus of enriched interstellar gas around the Sun is difficult
to explain without considering a link to the Gould Belt. The Belt is a
disk-like concentration of young stars, OB associations, neutral gas, molecular
clouds and dust in a plane intersecting the Galactic disk near the Sun, that is
inclined by approximately 20$^\circ$ with respect to the stars in
the Galactic disk \citep{tor00}. Its dominant stellar feature is a ring of OB
associations bounding this disk; the inner and outer edges of this ring are
approximated in Figure~\ref{krradplot} by elliptical annuli whose dimensions
and location are set by the \citet{per03} fit with an annulus width of about
80--100 pc. The Gould Belt's origins have generally been ascribed to a
combination of explosive events or the impact of a massive high velocity cloud
in the Solar neighborhood \citep{com94,pop97}. Regardless of the cause, the
Gould Belt includes OB associations formed in the last 30---60 Myr
\citep{tor00} at distances up to about 500 pc from the Sun, as well as gas
distributed throughout the region bounded by the OB associations in the
form of diffuse \ion{H}{1} and molecular clouds \citep{pop97,per03}. The
similarity between distances associated with the closer edge of the observed
annulus of abundance enhancements and those indicative of the outer reaches of
the Gould Belt is very suggestive, given that OB associations such as the ones
ringing Gould's Belt produce Type II supernovae. \ion{SN}{2} are important
production sites for krypton, oxygen, and copper that leave nitrogen
essentially unenhanced (e.g., \citealt{chi04}). In fact, the supernova rate for
the Gould Belt has been estimated at 3--5 times that of the local Galactic disk
\citep{gre00}. The observed outward flow of gas away from the Belt's central
regions \citep{per03} is consistent with both an explosive origin for the Belt
or the impact of a high velocity cloud \citep{com94}. The flow is also patchy,
which is consistent with the pattern of abundance enhancements identified in
this paper. Since the Belt was formed only a few tens of millions of years ago,
there has not been sufficient time for mixing to erase any abundance variations
introduced into interstellar gas by recent supernovae in the Belt.

It is worthwhile to note that analyses of abundances based on weak UV
absorption lines can be influenced by the selection effect alluded to in the
last few paragraphs. If the databases are restricted only to spectra with
absorption features strong enough to be detected but weak enough to avoid
concerns of unresolved saturation, and also to observations of stars bright
enough for the exposure time to be relatively short, then the analyses will
tend to be limited in the range of total hydrogen column density being sampled.
Consequently, as longer sight lines are included, more rarefied gas is being
probed. This tends to restrict the study of depletion variations to the local
ISM. In order to fully explore current Galactic elemental abundances and
depletion levels with a revitalized STIS and new Cosmic Origins Spectrograph,
as well as search for features such as the spatial abundance pattern noted in
this paper or other effects at large distances, it will be important to make
long exposure time observations of distant stars an important part of the
analysis.

\section{Conclusions}
This paper has combined new observations from an {\it HST} observing program
abbreviated by the STIS power failure with Archival detections and previously
published measurements of \ion{Kr}{1} $\lambda$1236 to compile a database of
50 Kr/H abundance ratios for sight lines up to 5.9 kpc in length and with
average hydrogen densities from less than 0.10 cm$^{-3}$ to more than 10.0
cm$^{-3}$. Collectively, the data strengthen earlier evidence that krypton is
undepleted in the ISM and that the Kr/H ratio is remarkably uniform within 750
pc of the Sun. Furthermore, these two properties underline the significance of
an apparent enhancement of 0.10 dex in krypton abundance for sight lines of
length 750---2500 pc. This abundance elevation is matched by enhanced O/H
abundance ratios for the same sight lines, and similar distance dependencies in
both gas-phase oxygen and copper abundances among other sight lines. However,
N/O abundance ratios are reduced by a factor of two thirds in the same annulus.
These disparate pieces of evidence suggest a nucleosynthetic origin for the
abundance variations, while their spatial extent is comparable to the size of
the Gould Belt. It is possible that the noted elemental abundance variations
for Kr, O, Cu, and N arise out of star formation engendered as the Gould Belt
was formed. Given the spatial extent of D/H variations, the Gould Belt may be
at the bottom of this trend as well.

\acknowledgments
We would like to thank Paul Cartledge for advice on statistical tests and the
anonymous referee for their comments, which have significantly improved the
quality of this paper. Support for this work was provided by the Space
Telescope Science Institute through grants to Louisiana State University and
Northwestern University (both GO9855 and AR10643). This research has made use
of the SIMBAD database, operated at CDS (Strasborg France).

{\it Facilities:} \facility{FUSE (LWRS,MDRS,HIRS)}, \facility{HST (STIS)}.

\clearpage

\clearpage

\begin{deluxetable}{lccccccc}
\tablecaption{Summary of Observations \label{datalist}}
\tablewidth{0pt}
\tablehead{
&\colhead{\it HST}&\colhead{\it STIS}&&\colhead{Time}&\colhead{{\it FUSE}}&
&\colhead{Time}\\
Sight Line&\colhead{Program}&\colhead{Data Set}&\colhead{Date}&\colhead{(s)}&
\colhead{Data Set}&\colhead{Date}&\colhead{(s)}\\
}
\startdata
HD 23478 & GO9505 & O6LJ01020 & 2002 Aug 20 & 2945 & A1200202000 & 2001 Jan 02 &
 1871   \\
HD 24190 & GO9505 & O6LJ02020 & 2003 Mar 28 & 2940 & A1200101000 & 2000 Sep 11 &
 4721   \\
HD 40893 & GO9855 & O8NA02010 & 2004 Feb 20 & 2162 & P2160101000 & 2001 Oct 14 &
 7138   \\
         &        & O8NA02020 & 2004 Feb 20 & 7902 & \nodata     & \nodata     &
\nodata \\
HD 69106 & GO8484 & O5LH03010 & 2000 Jun 08 &  660 & P1022301000 & 2001 Apr 04 &
   58   \\
         &        & O5LH03020 & 2000 Jun 08 &  625 & P1022302000 & 2002 Dec 31 &
  404   \\
HD 94454 & GO9505 & O6LJ0H010 & 2002 Dec 17 & 1780 & A1200501000 & 2000 Apr 04 &
 7622   \\
HD 99857 & GO8043 & O54301010 & 1999 Feb 20 & 1307 & P1024501000 & 2000 Feb 05 &
 4267   \\
         &        & O54301020 & 1999 Feb 20 & 1307 & \nodata     & \nodata     &
\nodata \\
         &        & O54301030 & 1999 Feb 20 & 1307 & \nodata     & \nodata     &
\nodata \\
HD 99872 & GO9505 & O6LJ0I020 & 2002 May 06 & 1890 & A1200606000 & 2000 Apr 09 &
 3270   \\
HD 102065& GO7509 & O4O001010 & 1998 Mar 28 & 7679 & Q1010101000 & 2000 May 28 &
 6793   \\
HD 104705& GO7270 & O57R01010 & 1998 Dec 24 & 2400 & P1025701000 & 2000 Feb 05 &
 4482   \\
HD 108639& GO9505 & O6LJ0A020 & 2002 Oct 19 & 1860 & A1201301000 & 2000 Mar 20 &
 4980   \\
HD 112999& GO9505 & O6LJ0C010 & 2002 Oct 21 & 1620 & A1202001000 & 2000 May 15 &
 7544   \\
         &        & O6LJ0C020 & 2002 Oct 21 & 1650 & \nodata     & \nodata     &
\nodata \\
HD 114886& GO9505 & O6LJ0D010 & 2003 Apr 03 & 1560 & A1201801000 & 2000 Apr 06 &
 4450   \\
         &        & O6LJ0D020 & 2003 Apr 03 & 1590 & \nodata     & \nodata     &
\nodata \\
HD 115071& GO9505 & O6LJ0E010 & 2003 Mar 31 & 1890 & P1025901000 & 2000 May 14 &
 5796   \\
         &        & O6LJ0E020 & 2003 Mar 31 & 1890 & \nodata     & \nodata     &
\nodata \\
HD 116852& GO8241 & O5C01C010 & 2000 Jun 25 &  360 & P1013801000 & 2000 May 27 &
 7212   \\
         & GO9855 & O8NA03010 & 2003 Aug 08 & 2522 & \nodata     & \nodata     &
\nodata \\
         &        & O8NA03020 & 2003 Aug 08 & 3200 & \nodata     & \nodata     &
\nodata \\
HD 122879& GO8241 & O5C037010 & 2000 May 08 &  360 & B0710501000 & 2002 Mar 03 &
 1532   \\
         & GO8484 & O5LH07010 & 2000 Aug 29 &  708 & \nodata     & \nodata     &
\nodata \\
         &        & O5LH07020 & 2000 Aug 29 &  708 & \nodata     & \nodata     &
\nodata \\
HD 124314& GO8043 & O54307010 & 1999 Apr 10 & 1466 & P1026201000 & 2000 Mar 22 &
 4382   \\
HD 137595& GO9505 & O6LJ03010 & 2002 Aug 21 & 1200 & A1201213000 & 2000 Jul 10 &
 5715   \\
HD 144965& GO9505 & O6LJ05010 & 2003 Feb 01 & 1510 & A1201601000 & 2000 Mar 21 &
 1195   \\
         &\nodata & \nodata   & \nodata    &\nodata& A1201602000 & 2001 Aug 08 &
 7220   \\
HD 147165& GO5891 & Z34R0407T & 1996 Mar 23 & 108.8& \nodata     & \nodata     &
\nodata \\
HD 147683& GO9505 & O6LJ06020 & 2003 Mar 27 & 2940 & A1200909000 & 2000 Jul 10 &
 7367   \\
HD 151805& GO9434 & O6LZ63010 & 2003 Apr 03 & 1200 & P1026602001 & 2000 Jul 14 &
 2741   \\
HD 152590& GO8241 & O5C08P010 & 2000 Mar 03 & 1440 & B0710601000 & 2001 Jul 08 &
 1389   \\
         & GO9855 & O8NA04010 & 2003 Sep 17 & 2240 & B0710602000 & 2001 Aug 12 &
 2773   \\
         &        & O8NA04020 & 2003 Sep 17 & 8076 & \nodata     & \nodata     &
\nodata \\
HD 165246& GO9855 & O8NA05010 & 2003 Aug 13 & 2158 & P1050301000 & 2000 Sep 01 &
 7782   \\
         &        & O8NA05020 & 2003 Aug 13 & 2626 & \nodata     & \nodata     &
\nodata \\
HD 177989& GO7270 & O57R03010 & 1999 May 28 & 1660 & P1017101000 & 2000 Aug 28 &
10289   \\
         &        & O57R03020 & 1999 May 28 & 2897 & \nodata     & \nodata     &
\nodata \\
HD 203374& GO848  & O5LH08010 & 2000 Mar 23 & 1325 & B0300101000 & 2001 Aug 02 &
10170   \\
         &        & O5LH08020 & 2000 Mar 23 & 1325 & B0300102000 & 2001 Aug 02 &
 8483   \\
         &        & O5LH08030 & 2000 Mar 23 & 1325 & \nodata     & \nodata     &
\nodata \\
HD 206267& GO8484 & O5LH09010 & 2000 Mar 24 &  711 & B0460301000 & 2002 Jun 30 &
 4869   \\
         &        & O5LH09020 & 2000 Mar 24 &  711 & \nodata     & \nodata     &
\nodata \\
HD 208947& GO8484 & O5LH0A010 & 2000 Jul 14 &  714 & \nodata     & \nodata     &
\nodata \\
         &        & O5LH0A020 & 2000 Jul 14 &  714 & \nodata     & \nodata     &
\nodata \\
HD 209339& GO8484 & O5LH0B010 & 2000 May 18 &  708 & B0300501000 & 2001 Jul 21 &
 2044   \\
         &        & O5LH0B020 & 2000 May 18 &  708 & \nodata     & \nodata     &
\nodata \\
HD 224151& GO8043 & O54308010 & 1999 Feb 18 & 1496 & P1224101000 & 2000 Aug 11 &
 6023   \\
         &\nodata & \nodata   & \nodata    &\nodata& S3040201000 & 1999 Dec 09 &
 7092   \\
         &\nodata & \nodata   & \nodata    &\nodata& S3040202000 & 2000 Aug 13 &
 9202   \\
HD 303308& GO7301 & O4QX04010 & 1998 Mar 19 & 2220 & P1221601000 & 2000 May 25 &
 6105   \\
         &\nodata & \nodata   & \nodata    &\nodata& P1221602000 & 2000 May 27 &
 7692   \\
\enddata
\end{deluxetable}

\clearpage

\begin{deluxetable}{lr@{ }lr@{ }lr@{ }lr@{ }lr@{ }lr@{ }llcr@{.}lc}
\rotate
\tabletypesize{\footnotesize}
\tablecaption{New Krypton and Hydrogen Sight Line Properties \label{KrHdata}}
\tablewidth{0pt}
\tablehead{
&\multicolumn{2}{c}{$W_{\lambda1235}$}&\multicolumn{2}{c}{log$_{10}$[$N$(Kr)]}&
\multicolumn{2}{c}{log$_{10}$[$N$(H)]}&
\multicolumn{2}{c}{log$_{10}$[$N$(\ion{H}{1})]}&
\multicolumn{2}{c}{log$_{10}$[$N$(H$_2$)]}&
\multicolumn{2}{c}{ }&\colhead{Spectral}&\colhead{$d_\ast$}&
\multicolumn{2}{c}{log$_{10}{\langle}n_{\rm H}\rangle$}&\colhead{ }\\
\colhead{Star}&\multicolumn{2}{c}{(m\AA)}&\multicolumn{2}{c}{[cm$^{-2}$]}&
\multicolumn{2}{c}{[cm$^{-2}$]}&\multicolumn{2}{c}{[cm$^{-2}$]}&
\multicolumn{2}{c}{[cm$^{-2}$]}&\multicolumn{2}{c}{log$_{10}$[Kr/H]}&
\colhead{Type}&\colhead{(kpc)}&\multicolumn{2}{c}{[cm$^{-3}$]}&
\colhead{log$_{10}f$(H$_2$)}\\
}
\startdata
HD 23478\tablenotemark{a}&4.16&(0.33)&12.23&(0.05)&21.21&(0.04)&21.01&(0.06)&
20.48 & (0.07) & $-$8.98 &$^{+0.06}_{-0.07}$& B3IV   &0.32&    0&22 & $-$0.43 \\
HD 24190      & 5.87 & (0.72) & 12.35 & (0.06) & 21.30 & (0.05) & 21.18 &(0.06)&
20.38 & (0.07) & $-$8.95 &$^{+0.08}_{-0.09}$& B2V    &0.32&    0&31 & $-$0.62 \\
HD 40893      & 9.52 & (1.35) & 12.56 & (0.07) & 21.54 & (0.05) & 21.45 &(0.09)&
20.49 & (0.07) & $-$8.98 &$^{+0.08}_{-0.10}$& B0IV   &2.80& $-$0&40 & $-$0.75 \\
HD 69106      & 4.06 & (0.76) & 12.19 & (0.09) & 21.09 & (0.06) & 21.06 &(0.06)&
19.64 & (0.07) & $-$8.90 &$^{+0.11}_{-0.14}$& B0.5II &1.40& $-$0&55 & $-$1.15 \\
HD 94454\tablenotemark{b}&5.37&(0.85)&12.36&(0.08)&\multicolumn{2}{c}{\nodata}&
\multicolumn{2}{c}{\nodata}&20.70&(0.07)&\multicolumn{2}{c}{\nodata}&
B8III   & 0.33 &\multicolumn{2}{c}{\nodata}&\nodata\\
HD 99857      & 5.47 & (0.94) & 12.34 & (0.08) & 21.36 & (0.05) & 21.28 &(0.06)&
20.29 & (0.07) & $-$9.02 &$^{+0.09}_{-0.12}$& B1Ib   &3.06& $-$0&61 & $-$0.77 \\
HD 99872\tablenotemark{b}&4.27&(0.47)&12.25&(0.07)&\multicolumn{2}{c}{\nodata}&
\multicolumn{2}{c}{\nodata}&20.55&(0.11)&\multicolumn{2}{c}{\nodata}&
B3V     & 0.23 &\multicolumn{2}{c}{\nodata}&\nodata\\
HD 102065\tablenotemark{b}&3.76&(0.52)&12.16&(0.11)&
\multicolumn{2}{c}{\nodata}&\multicolumn{2}{c}{\nodata}&20.47&(0.08)&
\multicolumn{2}{c}{\nodata}&B2V   & 0.17 &\multicolumn{2}{c}{\nodata}&\nodata\\
HD 104705     & 2.99 & (0.54) & 12.05 & (0.09) & 21.20 & (0.05) & 21.15 &(0.06)&
19.94 & (0.06) & $-$9.15 &$^{+0.10}_{-0.13}$&B0III/IV&3.90& $-$0&88 & $-$0.96 \\
HD 108639     & 6.72 & (1.34) & 12.40 & (0.10) & 21.38 & (0.08) & 21.35 &(0.09)&
19.95 & (0.07) & $-$8.98 &$^{+0.12}_{-0.17}$& B1III  &0.11&    0&84 & $-$1.13 \\
HD 112999\tablenotemark{b}&4.50&(0.65)&12.27&(0.09)&\multicolumn{2}{c}{\nodata}&
\multicolumn{2}{c}{\nodata}&19.99&(0.07)&\multicolumn{2}{c}{\nodata}&
B6IIIn & 0.34 &\multicolumn{2}{c}{\nodata}&\nodata\\
HD 114886     & 8.31 & (2.01) & 12.43 & (0.14) & 21.40 & (0.05) & 21.34 &(0.06)&
20.23 & (0.07) & $-$8.97 &$^{+0.15}_{-0.22}$& O9V    &0.38&    0&32 & $-$0.87 \\
HD 115071     &10.77 & (1.56) & 12.64 & (0.07) & 21.50 & (0.05) & 21.36 &(0.06)&
20.63 & (0.07) & $-$8.86 &$^{+0.08}_{-0.10}$& O9.5V  &1.64& $-$0&20 & $-$0.57 \\
HD 116852     & 3.02 & (0.38) & 12.06 & (0.06) & 21.03 & (0.06) & 20.98 &(0.06)&
19.79 & (0.11) & $-$8.97 &$^{+0.08}_{-0.10}$& O9III  &4.80& $-$1&14 & $-$0.94 \\
HD 122879     & 9.27 & (1.26) & 12.60 & (0.13) & 21.35 & (0.10) & 21.26 &(0.12)&
20.31 & (0.08) & $-$8.75 &$^{+0.16}_{-0.25}$& B0Ia   &2.10& $-$0&46 & $-$0.74 \\
HD 124314     &12.72 & (2.50) & 12.70 & (0.09) & 21.47 & (0.05) & 21.39 &(0.06)&
20.40 & (0.07) & $-$8.77 &$^{+0.10}_{-0.13}$& O7     &1.15& $-$0&08 & $-$0.77 \\
HD 137595\tablenotemark{a}&5.50&(0.63)&12.34&(0.07)&21.24&(0.04)&21.00&(0.06)&
20.56 & (0.06) & $-$8.90 &$^{+0.08}_{-0.10}$& B3Vn   &0.83& $-$0&17 & $-$0.38 \\
HD 144965\tablenotemark{a}&4.79&(0.70)&12.30&(0.08)&21.37&(0.05)&21.07&(0.06)&
20.77 & (0.07) & $-$9.07 &$^{+0.09}_{-0.12}$& B3Vne  &0.68&    0&05 & $-$0.30 \\
HD 147165     & 6.71 & (0.41) & 12.45 & (0.05) & 21.40 & (0.08) & 21.38 &(0.08)&
19.79 & (0.06) & $-$8.95 &$^{+0.09}_{-0.12}$& B1III  &0.15&    0&75 & $-$1.31 \\
HD 147683\tablenotemark{a}& 8.32 & (1.03) & 12.52 & (0.06) & 21.55 & (0.07) &
21.41 &(0.08)& 20.68 & (0.12) & $-$9.03 &$^{+0.09}_{-0.11}$& B4V    & 0.59 &
   0&29 & $-$0.57 \\
HD 151805     & 6.54 & (1.30) & 12.41 & (0.08) & 21.41 & (0.05) & 21.32 &(0.06)&
20.36 & (0.07) & $-$9.00 &$^{+0.13}_{-0.18}$& B1Ib   &5.91& $-$0&85 & $-$0.75 \\
HD 152590     & 8.49 & (1.05) & 12.54 & (0.05) & 21.47 & (0.05) & 21.37 &(0.06)&
20.47 & (0.07) & $-$8.93 &$^{+0.07}_{-0.08}$& O7.5V  &1.80& $-$0&30 & $-$0.70 \\
HD 165246     & 7.95 & (0.78) & 12.53 & (0.06) & 21.46 & (0.06) & 21.41 &(0.07)&
20.15 & (0.08) & $-$8.93 &$^{+0.08}_{-0.10}$& O9III  &1.59& $-$0&23 & $-$1.01 \\
HD 177989     & 4.70 & (0.72) & 12.25 & (0.07) & 21.09 & (0.05) & 20.98 &(0.06)&
20.16 & (0.08) & $-$8.84 &$^{+0.08}_{-0.10}$& B2II   &4.91& $-$1&09 & $-$0.63 \\
HD 203374     & 9.36 & (0.95) & 12.56 & (0.06) & 21.38 & (0.06) & 21.20 &(0.08)&
20.60 & (0.09) & $-$8.82 &$^{+0.08}_{-0.10}$& B0IVpe &0.62&    0&10 & $-$0.48 \\
HD 206267     &11.16 & (1.47) & 12.65 & (0.10) & 21.46 & (0.04) & 21.20 &(0.06)&
20.81 & (0.05) & $-$8.81 &$^{+0.11}_{-0.14}$& O6e    &0.59&    0&20 & $-$0.35 \\
HD 208947\tablenotemark{c}&2.53&(0.40)&11.99&(0.08)&21.10&(0.06)&21.10&(0.06)&
\multicolumn{2}{c}{\nodata}& $-$9.11 &$^{+0.09}_{-0.12}$& B2V  & 0.50 &
$\geq-$0&08 & \nodata\\
HD 209339     & 5.81 & (0.55) & 12.35 & (0.06) & 21.26 & (0.06) & 21.17 &(0.07)&
20.24 & (0.08) & $-$8.91 &$^{+0.08}_{-0.10}$& B0IV   &1.00& $-$0&23 & $-$0.72 \\
HD 224151     & 9.87 & (2.82) & 12.58 & (0.12) & 21.43 & (0.05) & 21.30 &(0.06)&
20.53 & (0.06) & $-$8.85 &$^{+0.13}_{-0.18}$& B0.5II &1.10& $-$0&10 & $-$0.60 \\
HD 303308     & 5.36 & (0.96) & 12.32 & (0.09) & 21.46 & (0.06) & 21.41 &(0.06)&
20.18 & (0.12) & $-$9.14 &$^{+0.11}_{-0.14}$& O3V    &2.67& $-$0&46 & $-$0.98 \\
\enddata
\tablenotetext{a}{The spectral type for the star associated with this sight
line suggests that the interstellar Ly-$\alpha$ profile is potentially
contaminated by stellar neutral hydrogen. These Kr/H ratios are used in the
analysis with caution.}
\tablenotetext{b}{The $N$(H), $N$(\ion{H}{1}), ${\langle}n_{\rm H}\rangle$, and
$f$(H$_2$) values are undetermined, due to obvious contamination of the
interstellar Ly-$\alpha$ line by stellar material. These sight lines are
omitted from the Kr/H ratio analysis.}
\tablenotetext{c}{No H$_2$ data are available; this sight line is omitted from
the Kr/H ratio analysis.}
\end{deluxetable}

\clearpage

\begin{deluxetable}{r@{$-$}lccccccccccc}
\rotate
\tabletypesize{\footnotesize} 
\tablecaption{Interstellar Gas-phase Abundance Weighted Means \label{aratmeans}}
\tablewidth{0pt}
\tablehead{
\multicolumn{2}{c}{Distance}&&&&&&&&&&&\\
\multicolumn{2}{c}{Range}&&\colhead{Error}&\colhead{Scatter}&\colhead{Sample}&
&\colhead{Error}&\colhead{Scatter}&&\colhead{Error}&\colhead{Scatter}&
\colhead{Sample}\\
\multicolumn{2}{c}{(kpc)}&\colhead{log$_{10}$[Kr/H]$_{ISM}$}&\colhead{(dex)}&
\colhead{(dex)}&\colhead{Size}&\colhead{log$_{10}$[O/H]$_{ISM}$}&
\colhead{(dex)}&\colhead{(dex)}&\colhead{log$_{10}$[O/Kr]$_{ISM}$}&
\colhead{(dex)}&\colhead{(dex)}&\colhead{Size}\\
}
\startdata
0   &0.75&$-9.02$&0.02&0.05&27&$-3.48$&0.03&0.04&5.59&0.04&0.07&6 \\
0.55&2.50&$-8.91$&0.02&0.08&16&$-3.40$&0.02&0.08&5.53&0.03&0.06&11\\
2.50&6.00&$-9.03$&0.04&0.10&7 &$-3.47$&0.03&0.10&5.56&0.04&0.04&7 \\
\multicolumn{2}{c}{Near+Far}&$-9.02$&0.02&0.06&34&$-3.48$&0.02&0.08&5.57&0.03&0.05&13\\
\enddata
\end{deluxetable}

\clearpage

\begin{deluxetable}{lr@{ }lr@{ }lr@{ }lr@{ }lr@{ }lr@{ }lc}
\rotate
\tabletypesize{\footnotesize}
\tablecaption{Oxygen Abundances \label{OHdata}}
\tablewidth{0pt}
\tablehead{
&\multicolumn{2}{c}{$W_{\lambda1356}$}&\multicolumn{2}{c}{log$_{10}$[$N$(O)]}&
\multicolumn{2}{c}{log$_{10}$[$N$(Kr)]}&
\multicolumn{2}{c}{log$_{10}$[$N$(H)]}&
\multicolumn{2}{c}{ }&\multicolumn{2}{c}{ }&\colhead{$d_\ast$}\\
\colhead{Star}&\multicolumn{2}{c}{(m\AA)}&
\multicolumn{2}{c}{[cm$^{-2}$]}&\multicolumn{2}{c}{[cm$^{-2}$]}&
\multicolumn{2}{c}{[cm$^{-2}$]}&\multicolumn{2}{c}{log$_{10}$[O/Kr]}&
\multicolumn{2}{c}{log$_{10}$[O/H]}&
\colhead{(kpc)}\\
}
\startdata
HD 23478\tablenotemark{a}&  9.58 & (0.57) & 17.79 & (0.09) & 12.23 & (0.05) &
21.21 & (0.04) & 5.56 &$^{+0.10}_{-0.13}$& $-$3.42 &$^{+0.10}_{-0.12}$&0.32\\
HD 24190       & 13.51 & (0.96) & 17.91 & (0.05) & 12.35 & (0.06) &
21.30 & (0.05) & 5.56 &$^{+0.08}_{-0.09}$& $-$3.39 &$^{+0.07}_{-0.08}$&0.32\\
HD 40893       & 19.94 & (1.06) & 18.07 & (0.03) & 12.56 & (0.07) &
21.54 & (0.05) & 5.51 &$^{+0.08}_{-0.09}$& $-$3.47 &$^{+0.06}_{-0.07}$&2.80\\
HD 69106       &  8.34 & (0.67) & 17.69 & (0.05) & 12.19 & (0.09) &
21.09 & (0.06) & 5.50 &$^{+0.10}_{-0.13}$& $-$3.40 &$^{+0.08}_{-0.09}$&1.40\\
HD 94454\tablenotemark{b}& 10.12 & (0.80) & 17.83 & (0.04) & 12.36 & (0.08) &
\multicolumn{2}{c}{\nodata}& 5.47 &$^{+0.09}_{-0.11}$&
\multicolumn{2}{c}{\nodata}& 0.33\\
HD 99857       & 13.32 & (0.89) & 17.92 & (0.04) & 12.34 & (0.08) &
21.36 & (0.05) & 5.58 &$^{+0.09}_{-0.11}$& $-$3.44 &$^{+0.06}_{-0.07}$&3.06\\
HD 99872\tablenotemark{b}&  9.10 & (0.72) & 17.78 & (0.05) & 12.25 & (0.07) &
\multicolumn{2}{c}{\nodata}& 5.53 &$^{+0.08}_{-0.10}$&
\multicolumn{2}{c}{\nodata}& 0.23\\
HD 102065\tablenotemark{b}& 7.63 & (0.51) & 17.70 & (0.07) & 12.16 & (0.11) &
\multicolumn{2}{c}{\nodata}& 5.54 &$^{+0.13}_{-0.18}$&
\multicolumn{2}{c}{\nodata}& 0.17\\
HD 104705\tablenotemark{d}& 7.82 & (0.47) & 17.67 & (0.06) & 12.05 & (0.09) &
21.20 & (0.05) & 5.62 &$^{+0.10}_{-0.12}$& $-$3.53 &$^{+0.06}_{-0.07}$&3.90\\
HD 108639      & 20.81 & (1.79) & 18.09 & (0.06) & 12.40 & (0.10) &
21.38 & (0.08) & 5.69 &$^{+0.11}_{-0.15}$& $-$3.29 &$^{+0.10}_{-0.13}$&0.11\\
HD 112999\tablenotemark{b}& 9.01 & (1.04) & 17.75 & (0.07) & 12.27 & (0.09) &
\multicolumn{2}{c}{\nodata}& 5.48 &$^{+0.11}_{-0.15}$&
\multicolumn{2}{c}{\nodata}& 0.34\\
HD 114886      & 15.61 & (1.67) & 17.95 & (0.06) & 12.43 & (0.14) &
21.40 & (0.05) & 5.52 &$^{+0.15}_{-0.23}$& $-$3.45 &$^{+0.08}_{-0.09}$&0.38\\
HD 115071      & 26.12 & (1.53) & 18.20 & (0.03) & 12.64 & (0.07) &
21.50 & (0.05) & 5.56 &$^{+0.08}_{-0.09}$& $-$3.30 &$^{+0.06}_{-0.07}$&1.64\\
HD 116852      &  6.41 & (0.65) & 17.56 & (0.04) & 12.06 & (0.06) &
21.03 & (0.06) & 5.50 &$^{+0.07}_{-0.08}$& $-$3.47 &$^{+0.07}_{-0.08}$&4.80\\
HD 122879      & 22.39 & (2.07) & 18.10 & (0.04) & 12.60 & (0.13) &
21.35 & (0.10) & 5.50 &$^{+0.13}_{-0.20}$& $-$3.25 &$^{+0.11}_{-0.14}$&2.10\\
HD 124314      & 24.32 & (1.53) & 18.17 & (0.04) & 12.70 & (0.09) &
21.47 & (0.05) & 5.47 &$^{+0.10}_{-0.12}$& $-$3.30 &$^{+0.06}_{-0.07}$&1.15\\
HD 137595\tablenotemark{a}& 9.15 & (0.78) & 17.76 & (0.04) & 12.34 & (0.07) &
21.24 & (0.04) & 5.42 &$^{+0.08}_{-0.10}$& $-$3.48 &$^{+0.06}_{-0.06}$&0.83\\
HD 144965\tablenotemark{a}& 8.74 & (0.68) & 17.80 & (0.06) & 12.30 & (0.08) &
21.37 & (0.05) & 5.50 &$^{+0.08}_{-0.10}$& $-$3.57 &$^{+0.08}_{-0.09}$&0.68\\
HD 147165      & 14.62 & (0.58) & 17.97 & (0.10) & 12.45 & (0.05) &
21.40 & (0.08) & 5.52 &$^{+0.11}_{-0.15}$& $-$3.43 &$^{+0.12}_{-0.17}$&0.15\\
HD 147683\tablenotemark{a}&15.06 & (0.90) & 18.00 & (0.05) & 12.52 & (0.06) &
21.55 & (0.07) & 5.48 &$^{+0.08}_{-0.09}$& $-$3.55 &$^{+0.08}_{-0.10}$&0.59\\
HD 151805\tablenotemark{d}&16.57 & (1.58) & 17.98 & (0.07) & 12.41 & (0.08) &
21.41 & (0.05) & 5.57 &$^{+0.13}_{-0.20}$& $-$3.43 &$^{+0.08}_{-0.10}$&5.91\\
HD 152590      & 16.80 & (0.85) & 18.01 & (0.03) & 12.54 & (0.05) &
21.47 & (0.05) & 5.47 &$^{+0.06}_{-0.07}$& $-$3.46 &$^{+0.06}_{-0.07}$&1.80\\
HD 165246      & 15.63 & (1.20) & 18.00 & (0.04) & 12.53 & (0.06) &
21.46 & (0.06) & 5.47 &$^{+0.07}_{-0.08}$& $-$3.46 &$^{+0.06}_{-0.07}$&1.59\\
HD 177989      & 11.92 & (0.89) & 17.83 & (0.04) & 12.25 & (0.07) &
21.09 & (0.05) & 5.58 &$^{+0.08}_{-0.10}$& $-$3.26 &$^{+0.06}_{-0.07}$&4.91\\
HD 203374      & 17.36 & (0.84) & 18.02 & (0.04) & 12.56 & (0.06) &
21.38 & (0.06) & 5.46 &$^{+0.07}_{-0.08}$& $-$3.36 &$^{+0.07}_{-0.08}$&0.62\\
HD 206267      & 19.59 & (0.82) & 18.11 & (0.03) & 12.65 & (0.10) &
21.46 & (0.04) & 5.46 &$^{+0.10}_{-0.14}$& $-$3.35 &$^{+0.05}_{-0.06}$&0.59\\
HD 208947\tablenotemark{c}&10.11 & (1.02) & 17.68 & (0.03) & 11.99 & (0.08) &
21.10 & (0.06) & 5.69 &$^{+0.08}_{-0.10}$& $-$3.42 &$^{+0.06}_{-0.07}$&0.50\\
HD 209339      & 14.99 & (0.80) & 17.96 & (0.04) & 12.35 & (0.06) &
21.26 & (0.06) & 5.61 &$^{+0.07}_{-0.08}$& $-$3.30 &$^{+0.07}_{-0.08}$&1.00\\
HD 224151      & 20.28 & (2.27) & 18.08 & (0.06) & 12.58 & (0.12) &
21.43 & (0.05) & 5.50 &$^{+0.13}_{-0.19}$& $-$3.35 &$^{+0.08}_{-0.09}$&1.10\\
HD 303308\tablenotemark{d}&13.09 & (0.55) & 17.91 & (0.04) & 12.32 & (0.09) &
21.46 & (0.06) & 5.59 &$^{+0.10}_{-0.12}$& $-$3.55 &$^{+0.07}_{-0.08}$&2.67\\
\enddata
\tablenotetext{a}{Potential Ly-$\alpha$ contamination; hydrogen abundance
ratios used with caution.}
\tablenotetext{b}{There is apparent Ly-$\alpha$ contamination; hydrogen
abundance ratios are unavailable.}
\tablenotetext{c}{No H$_2$ data are available; ratios involving hydrogen refer
to \ion{H}{1} only.}
\tablenotetext{d}{The given oxygen abundance is tallied over radial velocities
consistent with the krypton profile; the measurement does not represent the
total oxygen abundance for this sight line.}
\end{deluxetable}

\end{document}